\documentclass[aps, pra, twocolumn,showpacs,preprintnumbers,amsmath,amssymb]{revtex4}

\newcommand{\ket}[1]{\mbox{$ | #1 \rangle $}}
\newcommand{\bra}[1]{\mbox{$ \langle #1 | $}}
\usepackage[dvips]{graphicx}

\begin{document}

\preprint{}

\title{Quantum secure communication scheme with W state}

\author{Jian Wang}

 \email{jwang@nudt.edu.cn}

\affiliation{School of Electronic Science and Engineering,
\\National University of Defense Technology, Changsha, 410073, China }
\author{Quan Zhang}
\affiliation{School of Electronic Science and Engineering,
\\National University of Defense Technology, Changsha, 410073, China }
\author{Chao-jing Tang}
\affiliation{School of Electronic Science and Engineering,
\\National University of Defense Technology, Changsha, 410073, China }


\begin{abstract}
Recently, Cao et al. proposed a new quantum secure direct
communication scheme using W state. In their scheme, the error rate
introduced by an eavesdropper who takes intercept-resend attack, is
only 8.3\%. Actually, their scheme is just a quantum key
distribution scheme because the communication parties first create a
shared key and then encrypt the secret message using one-time pad.
We then present a quantum secure communication scheme using
three-qubit W state. In our scheme, the error rate is raised to 25\%
and it is not necessary for the present scheme to use alternative
measurement or Bell basis measurement. We also show our scheme is
unconditionally secure.
\end{abstract}

\maketitle


%
%
Quantum key distribution (QKD) utilizes quantum effects to
distribute a secret key among legitimate parties
\cite{bb84,e91,bbm92}. Different to QKD, Quantum secure direct
communication (QSDC) is to transmit the secret message directly
without first establishing a key to encrypt them
\cite{si99,si00,beige}. QSDC can be used in some special
environments which has been shown by Bostr\"{o}em and Deng et
al.\cite{Bostrom,Deng}. Many researches have been carried out in
QSDC
\cite{si99,si00,beige,Bostrom,Deng,denglong,cai1,jwang,jwang1,jwang2,cw1,cw2,tg,zjz}.
These works can be divided into two kinds, one utilizes single
photons, the other utilizes entangled state. Deng et al. proposed a
QSDC scheme using batches of single photons which serves as one-time
pad \cite{denglong}. Cai et al. presented a deterministic secure
direct communication scheme using single qubit in a mixed state
\cite{cai1}. We proposed a QSDC scheme and a multiparty controlled
QSDC scheme using order rearrange of single photons \cite{jwang}.
Einstein-Podolsky-Rosen (EPR) pairs and Greenberger-Horne-Zeilinger
(GHZ) states are the main quantum channels exploited in the QSDC.
Deng et al. put forward a two-step QSDC protocol using
Einstein-Podolsky-Rosen (EPR) pairs \cite{Deng}. We presented a QSDC
scheme using EPR pairs and teleportation \cite{jwang1} and a
multiparty controlled QSDC scheme using Greenberger-Horne-Zeilinger
(GHZ) states \cite{jwang2}. Wang et al. proposed a QSDC scheme with
quantum superdense coding \cite{cw1} and a multi-step QSDC scheme
using GHZ state \cite{cw2}. Gao et al. and Zhang et al. each
presented a QSDC scheme using entanglement swapping \cite{tg,zjz}.

Entanglement is at the heart of quantum information processes. The
entanglement of three-qubit is classified by separable,
bi-separable, W, and GHZ state \cite{ckw,dvc,ghz}. W state has the
different physical properties from GHZ state \cite{jpok,wrz}. An
important characteristic of three-particle GHZ state is that loss of
any one of the qubits leaves the other two in a mixed state with
only classical correlations. W state is the 3-qubit state in which
each pair of qubits have the same and maximum amount of bipartite
entanglement. This feature makes the entanglement of the W state
maximally symmetrically robust against loss of any single qubit. The
GHZ class state cannot be transformed to the W class state under any
local operation and classical communication (LOCC).

In a recent Letter, Cao et al. proposed a novel QSDC scheme based on
a series of four-qubit W states and local Bell basis measurement
(hereafter called Cao's scheme) \cite{cao}. However, Cao's scheme is
not a genuine QSDC scheme. And if an eavesdropper, say Eve performs
intercept-resend attack on their scheme, the error rate introduced
by her is only 8.3\%. Then if the communication parties did not
detect the existence of Eve, all the secret messages will be stolen
by Eve. To improve the ability of eavesdropping check, we present a
quantum secure communication scheme using three-qubit W states. In
the present scheme, the error rate introduced by Eve can achieve
25\%. At the same time, the efficiency of the scheme is also
improved because the communication parties need only to perform
deterministic von Neumann measurement. Different to QKD, in our
scheme, the communication parties cannot establish a shared key
without the sender's measurement results. Only after obtaining the
sender's classical message could the receiver recover the sender's
secret message, which is different to QSDC in some sort. Therefore
we call the present scheme quantum secure communication scheme. The
security for the scheme is the same as that for BBM92 protocol
\cite{bbm92}, which is unconditional secure.

We first consider the intercept-resend attack in Cao's scheme. The
four-qubit symmetric W state can be written in different bases as
\begin{eqnarray}
\label{1}
\ket{W_4}&=&\frac{1}{2}(\ket{1000}+\ket{0100}+\ket{0010}+\ket{0001})_{1234}\nonumber\\
&=&\frac{1}{2}[(\ket{10}+\ket{01})\ket{00}+\ket{00}(\ket{10}+\ket{01})]\nonumber\\
&=&\frac{1}{2}[\ket{\psi^+}(\ket{\phi^+}+\ket{\phi^-})+(\ket{\phi^+}+\ket{\phi^-})\ket{\psi^+}]\nonumber\\
&=&\frac{1}{4}[\ket{++}(2\ket{++}+\ket{+-}+\ket{-+})\nonumber\\
& &-\ket{--}(2\ket{--}+\ket{+-}+\ket{-+})\nonumber\\
& &+\ket{+-}(\ket{++}-\ket{--})\nonumber\\
& &+\ket{-+}(\ket{++}-\ket{--})],
\end{eqnarray}
where $\ket{+}=\frac{1}{\sqrt{2}}(\ket{0}+\ket{1})$,
$\ket{-}=\frac{1}{\sqrt{2}}(\ket{0}-\ket{1})$. According to Cao's
scheme, Alice sends the $B$ sequence to Bob. Suppose Eve intercepts
the $B$ sequence and performs $Z$-basis (\ket{0}, \ket{1})
measurement on the two particles $P_i(3,4)$ in the $B$ sequence (In
this attack, Eve can also use Bell basis measurement). If Eve's
measurement result is \ket{00} (\ket{10} or \ket{01}), she resends
the particles 3, 4 in the state \ket{00} (\ket{\psi^+}) to Bob. In
Cao's scheme, Alice will choose randomly $Z$-basis, $X$-basis
(\ket{+}, \ket{-}) or Bell basis measurement to check eavesdropping.
Because of Eve's attack, the W state collapses to
\begin{eqnarray}
\label{2}
\ket{\Psi_1}&=&\frac{1}{\sqrt{2}}(\ket{10}+\ket{01})_{12}\ket{00}_{34}\nonumber\\
&=&\frac{1}{\sqrt{2}}\ket{\psi^+})_{12}(\ket{\phi^+}+\ket{\phi^-})_{34}\nonumber\\
&=&\frac{1}{\sqrt{2}}(\ket{++}-\ket{--})_{12}(\ket{++}\nonumber\\
& &+\ket{+-}+\ket{-+}+\ket{--})_{34}
\end{eqnarray}
or
\begin{eqnarray}
\label{3}
\ket{\Psi_1}&=&\frac{1}{\sqrt{2}}\ket{00}_{12}(\ket{10}+\ket{01})_{34}\nonumber\\
&=&\frac{1}{\sqrt{2}}(\ket{\phi^+}+\ket{\phi^-})_{12}\ket{\psi^+})_{34}\nonumber\\
&=&\frac{1}{\sqrt{2}}(\ket{++}+\ket{+-}+\ket{-+}+\ket{--})_{12}\nonumber\\
& &(\ket{++}-\ket{--})_{34}
\end{eqnarray}
each with probability 1/2. Obviously, Eve's attack will not
introduce any error if Alice and Bob perform $Z$-basis or Bell basis
measurement. If the two parties perform $X$-basis measurement, the
error rate introduced by Eve will be 1/4 according to the Eqs.
\ref{1}, \ref{2} and \ref{3}. Thus the total error rate is
$1/3\times1/4=0.083$. And if Alice utilizes this insecure quantum
channel to transmit her secret message, according to the process of
Cao's scheme, Eve will obtain all of Alice's secret messages because
Eve can make certain whether Alice's or Bob's measurement result is
$\ket{\psi^+}$ or $\ket{\phi^\pm}$ in this attack.

Suppose Alice encodes \ket{\psi^+}$\rightarrow$ 0,
\ket{\phi^\pm}$\rightarrow$ 1 and Bob encodes
\ket{\phi^\pm}$\rightarrow$ 0, \ket{\psi^+}$\rightarrow$ 1.
According to Cao's scheme, after the eavesdropping check, Alice and
Bob perform Bell basis measurements on their corresponding particles
and they then establish a shared key. The classical messages that
Alice sends to Bob are actually the data which Alice generated by
using their shared key to encrypt her secret messages. In other
words, Alice's outcome encoding $\oplus$ her secret message equals
to classical information, where $\oplus$ indicates modulo 2
addition. Bob can then recover the secret message by using his
outcome encoding $\oplus$ classical information, which is the same
as one-time pad.

We then present a quantum secure communication scheme using
three-qubit W state in order to improve the checking probability and
the efficiency for the scheme with W state. The details of our
scheme is as follows:

(S1) Alice prepares $N$ three-qubit W states each of which is
randomly in one of the two states
\begin{eqnarray}
\ket{\Phi_1}=\frac{1}{\sqrt{3}}(\ket{100}+\ket{010}+\ket{001})_{123},\\
\ket{\Phi_2}=\frac{1}{\sqrt{3}}(\ket{10+}+\ket{01+}+\ket{00-})_{123},
\end{eqnarray}
where 1, 2 and 3 represent the three particles of W state. We
denotes the ordered $N$ three-qubit W states with
[P$_1(1,2,3)$,P$_2(1,2,3)$,$\cdots$, P$_N(1,2,3)$] (hereafter called
$W$ sequence), where the subscript indicates the order of each
three-particle in the sequence. Alice takes the particles 1 and 2
from each state to form an ordered particle sequence [P$_1(1,2)$,
P$_2(1,2)$,$\cdots$, P$_N(1,2)$], called $A$ sequence. The remaining
partner particles compose $B$ sequence, [P$_1(3)$,
P$_2(3)$,$\cdots$, P$_N(3)$]. Alice selects randomly a sufficiently
large subset from the $W$ sequence for eavesdropping check, called
checking sequence ($C$ sequence). The remaining particles in the $W$
sequence form a message sequence ($M$ sequence).

(S2) Alice encodes her secret message on the $M$ sequence by
performing one of the two unitary operations
\begin{eqnarray}
& &I=\ket{0}\bra{0}+\ket{1}\bra{1},\nonumber\\
& &U=i\sigma_y=\ket{0}\bra{1}-\ket{1}\bra{0}.
\end{eqnarray}
on each of the particles 3 in the $M$ sequence. If her secret
message is ``0'' (``1''), Alice performs operation $I$ ($U$). The
operation $U$ flips the state in both $Z$-basis and $X$-basis, as
\begin{eqnarray}
& &U\ket{0}=-\ket{1}, U\ket{1}=\ket{0},\nonumber\\
& &U\ket{+}=\ket{-}, U\ket{-}=-\ket{+}.
\end{eqnarray}
Alice then sends the $B$ sequence to Bob.

(S3) After confirming Bob has received the $B$ sequence, Alice
announces publicly the initial states she prepared. If the initial
state is \ket{\Phi_1}, Bob has nothing to do. If the initial state
is \ket{\Phi_2}, he performs Hadamada operation on the particle 3 in
the $B$ sequence.

(S4) Alice publishes the position of the $C$ sequence. Both Alice
and Bob measure the sampling particles in the $Z$-basis. Alice let
Bob announce his measurement results. If Alice's result is \ket{10}
or \ket{01} (\ket{00}), Bob's result must be \ket{0} (\ket{1}). She
can then evaluate the error rate of the transmission of the $B$
sequence. If the error rate exceeds the threshold, they abort the
scheme. Otherwise, they continue to the next step.

(S5) Alice and Bob perform $Z$-basis measurements on their
corresponding particles in the $M$ sequence. Alice then publishes
her measurement results of the particles 2, 3 in the $M$ sequence.
Thus Bob can recover Alice's secret message, according to Alice's
result, as illustrated in Table 1.
\begin{table}[h]
\caption{The recovery of Alice's secret message }\label{Tab:one}
  \centering
    \begin{tabular}[b]{|c|c|c| c|} \hline
     Alice's result & Bob's result & secret message\\ \hline
      \ \ket{10} or \ket{01} & \ket{0}& 0 \\ \hline
       \ \ket{10} or \ket{01} & \ket{1} & 1\\ \hline
        \ \ket{00} & \ket{0} & 1\\ \hline
         \ \ket{00} & \ket{1} & 0\\ \hline
        \end{tabular}
\end{table}
Suppose Bob's result is $\ket{0}$. If Alice's measurement result of
particle 2, 3 in the $M$ sequence is \ket{00} (\ket{10} or
\ket{01}), they then conclude that Alice's secret message is ``1''
(``0'').

We now discuss the security for the present scheme. We first
consider the intercept-resend attack strategy. In this attack, Eve
intercepts the particles in the $B$ sequence and makes measurements
on them. Then she resends a particle sequence to Bob according to
her measurement results. In other words, the state of each particle
in the resend sequence is equal to her measurement result. Suppose
Eve measures the intercepted particle in $Z$-basis. If the initial
state is \ket{\Phi_2}, it collapses to
$\frac{1}{\sqrt{3}}$(\ket{10}+\ket{01}+\ket{00})\ket{0} or
$\frac{1}{\sqrt{3}}$(\ket{10}+\ket{01}-\ket{00})\ket{1} each with
probability 1/2. Thus the error rate introduced by Eve will be
$1/2\times1/3+1/2\times2/3=1/2$. If the initial state is
\ket{\Phi_1}, Eve's attack will not be detected. In this instance,
the total error rate is 25\%. Suppose Eve measures the intercepted
particle in the $X$-basis. Similarly, if the initial state is
\ket{\Phi_1}, the state will collapse to
$\frac{1}{\sqrt{3}}$(\ket{10}+\ket{01}+\ket{00})\ket{+} or
$\frac{1}{\sqrt{3}}$(\ket{10}+\ket{01}-\ket{00})\ket{-} each with
probability 1/2. According to the scheme, Bob will perform Hadamada
operation on the particle 3. Thus the error rate will also be 1/4.
Therefore, in the intercept-resend attack, the total error rate
introduced by Eve achieves 25\%.

We then consider the collective attack strategy. In this strategy,
Eve intercepts the particle P$_i(3)$ (i=1,2,$\cdots$,N) and uses it
and her own ancillary particle in the state $\ket{0}$ to do a CNOT
operation (the particle P$_i(3)$ is the controller, Eve's ancillary
particle is the target). Then Eve resends the particle P$_i(3)$ to
Bob. However, Eve cannot make certain the initial state which the
particle 3 belongs to. Suppose the initial state is \ket{\Phi_1},
Eve will attack successfully. But if the initial state is
\ket{\Phi_2}, the state is changed to
\begin{eqnarray}
\ket{\Phi_2'}&=&\frac{1}{\sqrt{6}}[(\ket{10}+\ket{01})(\ket{00}+\ket{11})\nonumber\\
& &+\ket{00}(\ket{00}-\ket{11})]_{123e},
\end{eqnarray}
where the subscript $e$ indicates Eve's ancillary particle.
According to the scheme, \ket{\Phi_2'} will collapse to
$\frac{1}{\sqrt{3}}$(\ket{10}+\ket{01}+\ket{00})\ket{00} or
$\frac{1}{\sqrt{3}}$(\ket{10}+\ket{01}-\ket{00})\ket{11} each with
probability 1/2. Similar to the analysis in the intercept-resend
attack, in this attack, the total error rate introduced by Eve is
also 25\%.

In fact, the security for the present scheme is based on
entanglement and random Hadamada operation. In the scheme, the
random Hadamada operation is equal to selecting $Z$-basis or
$X$-basis randomly to measure the transmitting particle. Note that
\begin{eqnarray}
\ket{\Phi_1}=\frac{1}{\sqrt{3}}[(\ket{10}+\ket{01})\ket{0}+\ket{00}\ket{1})]_{123},\\
\ket{\Phi_1}=\frac{1}{\sqrt{3}}[(\ket{10}+\ket{01})\ket{+}+\ket{00}\ket{-})]_{123}.
\end{eqnarray}
In this way, the security for the scheme is the same as that for
BBM92 protocol which is proved to be unconditionally secure. As we
described above, our scheme is also unconditionally secure.

So far we have presented a quantum secure communication scheme using
W state. The security for the scheme is equal to that for BBM92
protocol. Cao's scheme is not a genuine QSDC scheme because the
communication parties first establish a shared key and then encrypt
the sender's secret to the receiver. Strictly speaking, our scheme
is also not a QSDC scheme because only the sender's measurement
result has been published could the receiver recover the sender's
secret. Certainly, our scheme is not a QKD scheme because the
communication parties can not establish a shared key if the sender's
measurement result is not published. Therefore we call the present
scheme quantum secure communication scheme. In our scheme, all of
the W states are used to transmit the sender's secret message except
those chosen for checking eavesdropping. The communication parties
need only deterministic von Neumann measurement. In this way, the
present scheme is practical within today's technology.



\begin{acknowledgments}
This work is supported by the National Natural Science Foundation of
China under Grant No. 60472032.
\end{acknowledgments}

%
%

%
%

\begin{thebibliography}{99}
\bibitem{bb84} C. H. Bennett and G. Brassard, \textit{in Proceedings of IEEE
international Conference on Computers, Systems and signal
Processing, Bangalore, India} (IEEE, New York), pp. 175 - 179
(1984).
\bibitem{e91} A. K. Ekert, Phys. Rev. Lett. \textbf{67}, 661 (1991).
\bibitem{bbm92} C. H. Bennett, G. Brassard and N. D. Mermin, Phys. Rev. Lett.
\textbf{68}, 557 (1992).
\bibitem{si99} K. Shimizu and N. Imoto, Phys. Rev. A \textbf{60}, 157 (1999).
\bibitem{si00} K. Shimizu and N. Imoto, Phys. Rev. A \textbf{62}, 054303 (2000).
\bibitem{beige} A. Beige, B.-G. Englert, Ch. Kurtsiefer, and
H. Weinfurter, Acta Phys. Pol. A \textbf{101}, 357 (2002).
\bibitem{Bostrom} K. Bostr\"{o}em and T. Felbinger, Phys. Rev. Lett. \textbf{89}, 187902 (2002).
\bibitem{Deng} F. G. Deng, G. L. Long, and X. S. Liu, Phys. Rev. A \textbf{68}, 042317 (2003).
\bibitem{denglong} F. G. Deng and G. L. Long, Phys. Rev. A \textbf{69}, 052319 (2004).
\bibitem{cai1} Q. Y. Cai and B. W. Li, Chin. Phys. Lett. \textbf{21}, 601 (2004).
\bibitem{jwang} J. Wang, Q. Zhang and C. J. Tang, quant-ph/0603100.
\bibitem{jwang1} J. Wang, Q. Zhang and C. J. Tang, quant-ph/0511092.
\bibitem{jwang2} J. Wang, Q. Zhang and C. J. Tang, quant-ph/0602166.
\bibitem{cw1} C. Wang, F. G. Deng, Y. S. Li, X. S. Liu and G. L. Long, Phys. Rev. A \textbf{71}, 044305 (2005).
\bibitem{cw2} C. Wang, F. G. Deng and G. L. Long, Opt. Commun. \textbf{253}, 15 (2005).
\bibitem{tg} T. Gao, F. L. Yan and Z. X. Wang, quant-ph/0406083.
\bibitem{zjz} Z. J. Zhang and Z. X. Man, quant-ph/040321.
\bibitem{ckw} V. Coffman, J. Kundu and W. K. Wootters, Phys. Rev. A \textbf{61}, 052306 (2000).
\bibitem{dvc} W. D\"{u}r, G. Vidal and J. I. Cirac Phys. Rev. A \textbf{62}, 062314 (2000).
\bibitem{ghz} D. M. Greenberger, M. A. Horne and A. Zeilinger, Am. J. Phys \textbf{58}, 1131 (1990).
\bibitem{jpok} J. Joc, Y. J. Park, S. Oh and J. Kim, New J. Phys \textbf{5}, 136 (2003).
\bibitem{wrz} P. Walther, K. J. Resch and A. Zeilinger, Phys. Rev. Lett. \textbf{94}, 240501 (2005).
\bibitem{cao} H. J. Cao and H. S. Song, Chin. Phys. Lett. \textbf{23}, 290 (2006).
\end{thebibliography}
\end{document}